# Filling the citation gap: Measuring the multidimensional impact of the academic book at institutional level with PlumX

**Daniel Torres-Salinas[1], Nicolás Robinson-Garcia[2*] and Juan Gorraiz[3]**

[1] Universidad de Granada (EC3metrics & Medialab UGR) and Universidad de Navarra

[2] INGENIO (CSIC-UPV), Universitat Politècnica de València, Valencia (Spain)

[3] Vienna University Library, University of Vienna, Vienna (Austria)

* Corresponding author: elrobin@ingenio.upv.es

## Abstract

More than five years after their emergence, altmetrics are still seen as a promise to complement traditional citation-based indicators. However, no study has focused on their potential usefulness to capture the impact of scholarly books. While recent literature shows that citation indicators cannot fully capture the impact of books, other studies have suggested alternative indicators such as usage, publishers' prestige or library holdings. In this paper, we calculate 18 indicators which range from altmetrics to library holdings, views, downloads or citations to the production of monographs of a Spanish university using the bibliometric suite PlumX from EBSCO. The objective of the study is to adopt a multidimensional perspective on the analysis of books and understand the level of complementarity between these different indicators. Also, we compare the overview offered by this range of indicators when applied to monographs with the traditional bibliometric perspective focused on journal articles and citation impact. We observe a low presence of altmetric indicators for monographs, even lower than for journal articles and a predominance of library holdings, confirming this indicator as the most promising one towards the analysis of the impact of books.

## Introduction

There are increasing voices alerting on the uncertainty of the future of the scholarly monographs (e.g., Williams et al., 2009; Watkinson et al., 2016). Still, they remain the primary academic output in the arts, humanities and some social sciences (Nederhof, 2006; Huang & Chang, 2008). Indeed, the 2014 UK Research Excellence Framework (REF, 2014) reported that book submissions represented 55% for the humanities, 33% for the arts, and 22% for the social sciences of all submissions in these fields. On the other end, books represented about 0.5% of all submissions in science, engineering and medicine (Kousha et al., 2016). Despite this, books are still not fairly assessed in evaluation exercises. Until recently, books were absent in the main bibliometric databases, leading to a devaluation of monographs as a secondary scientific product. Due to the pressure exerted by national evaluation schemes, many researchers shift or have shifted from books to journal articles as their preferred dissemination channel (Research Information Network, 2009). Furthermore, almost all university rankings, including the ones based solely on bibliometric data - like the Leiden Ranking - ignore them even in disciplines where they play a crucial role. There are only few bibliometric analyses with evaluative purposes considering their importance in such fields, and all of them corroborate the important role of monographs and the significant contribution to citation analyses (e.g. Kousha & Thelwall, 2009; Gorraiz et al, 2016).

The assessment of the impact of monographs is nowadays a big challenge and a hot topic in the scientometric field. Citation analyses are an acceptable proxy for the measurement of the impact of research publications, but only for a subset of the scientific community, namely the ''publish or perish'' group and only of the impact reflected by documented scholarly communication. However, it is common knowledge that many disciplines address much broader audiences within the academic community and even beyond. Monographs can have educational or public interest value as well as research impact (Kousha & Thelwall, 2015) and they can aim to enrich culturally non-academic audiences (Small, 2013). In this context, new metrics and specifically usage metrics (Gorraiz et al., 2014b; Glänzel & Gorraiz, 2015) and altmetrics (Priem, 2014; Robinson-Garcia et al., 2014), have the potential to apply alternative evaluation methods that complement citation-based indicators. Gaining a much broader and more accurate picture of the impact of monographs.

The launch of the Book Citation Index (BKCI) in 2011 enabled and eased access to citation data for large collections of books. It opened the floor for a large amount of citation studies on the citation patterns, characteristics and peculiarities of books (e.g. Kousha et al, 2011; Leydesdorff & Felt, 2012; Gorraiz et al, 2013; Torres-Salinas et al, 2012; 2013; 2014a,b). Still, many shortcomings must be surpassed before being able to use citation data for evaluative purposes. Some of these shortcomings are due to coverage and technical issues of the data sources (Torres-Salinas et al., 2014a) while others are related with the design and conceptualization of the indicators (Chi, 2016). Additionally, other approaches have been suggested in the literature. Following, we mention the main ones:

a) **Library Catalog Analysis**. Based on the use of library holdings per book title. Here several approaches are presented. Torres-Salinas and Moed (2009) used the number of catalog entries per book title in the WorldCat catalog (Torres-Salinas and Moed, 2009). Linmans (2010) used library bindings (Linmans 2010), while White et al. (2009) even considered it as an indicator of perceived cultural benefit.

b) **Library loans**. Influenced partly by the Library Catalog Analysis method and by the work of Schlogl and Gorraiz (2006), Cabezas-Clavijo et al. (2013) suggest that library loans may be a potential proxy for measuring the use of books. However, problems related with data cleaning and missing data prevent from further expanding this methodology.

c) **Publishers' prestige**. Here we observe two approaches. Giménez-Toledo et al. (2012) elaborated a publishers' ranking based on experts' opinions. This approach has been implemented with different methodological variations in countries such as Spain, Denmark, Finland or Norway (Giménez-Toledo et al., 2016). Torres-Salinas et al. (2012) used a more traditional approach based on citations from the Book Citation Index to develop a set of indicators by publisher and field, simulating the Journal Citation Reports.

d) **Book reviews**. If we consider the book as the main unit of analysis (disregarding publishers or book chapters), book reviews, which are extensively used for informing on recently released books, could be used to quantify the value of books. Zuccala & van Leeuwen (2011) were the first ones to suggest such approach. Since, other studies have followed. For instance, Gorraiz et al. (2014a) suggested this methodology not as a substitute, but as a complement to surpass coverage limitations from the Book Citation Index. Another perspective is that where social platforms for books are used to retrieve users' opinions and reviews (Kousha & Thelwall 2015 a,b; Kousha et al., 2016; Zuccala et al., 2015).

As observed, in this wide variety of proposals, little consensus can be found on which is the best proxy of quality or impact to use. The unit of analysis differs depending on the proposal. While in some cases, they focus on books, in others the focus is on publishers or in book chapters. Also, with some exceptions (Zuccala et al., 2015), there is no conceptual analysis on the meaning of the different impact proxies used (i.e., citations, library holdings, etc.).

There is an increasing interest on the use of altmetrics for analysing scholarly impact (Priem, 2014) and many studies have been devoted towards analysing its potential and caveats (i.e., Costas, Zahedi, & Wouters, 2015; Haustein, 2016; Thelwall, Haustein, Larivière, & Sugimoto, 2013). Also, many commercial solutions are currently available to recollect social media mentions. So far, the main one being used in research is Altmetric.com. Altmetric.com is a company owned by Digital Science which recollects mentions from a wide variety of social media platform to scientific publications (Robinson-García, Torres-Salinas, Zahedi, & Costas, 2014). However, most of these studies are mainly focused on journal articles. Recently, new sources and collaborative projects have been launched to cover this gap. On the one hand, altmetric.com launched in collaboration with Springer, the Bookmetrix project (http://www.bookmetrix.com), in which altmetric indicators are provided at the book and book chapter level for all Springer publications.

Another commercial solution is the offered by EBSCO (and recently acquired by Elsevier); the PlumX suite. This platform includes books as well as journal articles and provides a wide range of indicators as well as altmetric indicators. This allows considering simultaneously many of these proxies and offer a multidimensional approach of the impact of monographs. Hence, comparing the results provided by each one, and identifying the more relevant indicators. Altmetric indicators have only been explored for books by focusing on specific altmetric indicators (Kousha & Thelwall, 2015). This study is a first explorative attempt to provide such a multidimensional approach. We analyse 18 different indicators related with citations, downloads, library holdings and altmetrics for a given university. The main goal is to study the relation of these different dimensions altogether to better comprehend how they complement each other and in which cases certain indicators may provide more or less information than others. Specifically, we pose the following objectives:

1.  Analyse the different dimensions present in PlumX based on the 18 indicators they provide and its potential usefulness for research evaluation purposes.

2.  Explore to what extent these indicators complement the information reported by traditional evaluations focused on journal output.

For this, we analyse the output of monographs of the University of Granada. This is the first study using PlumX to analyse the scholarly impact of monographs. It is also the first study adopting a multidimensional perspective for the assessment of the broad impact of books.

# Material and methods

This paper analyzes and compares a set of 18 indicators for a set of monographs. For this, we take as a sample a set of monographs published by the University of Granada between 2010 and 2016, and retrieved from the Andalusian Current Research Information System (CRIS). This allows working with a large data set where all scientific fields are represented. In this section, we describe the data collection process, the characteristics of the data sources employed and the indicators used. First, we describe our publication data collection in order to give an overview of the coverage by fields, how was data collected, etc. Next, we describe the data source employed to obtain the impact indicators: PlumX.

## Publication data collection

By the end of September, 2016, a data set of monographs published by the University of Granada between 2010 and 2016 was retrieved from the Andalusian CRIS (known by its Spanish acronym

SICA). For this, all records including an ISBN number and registered as books were identified and processed into a relational database. Figure 1 includes the annual distribution of monographs published during the period of study. A total of 2,957 books were retrieved, from which 24% were published in 2010. Since then we observe a declining trend on the production of books. This could be mainly since the information included in SICA is self-reported and it is not mandatory, hence this trend cannot be interpreted as a decline on the production of monographs.

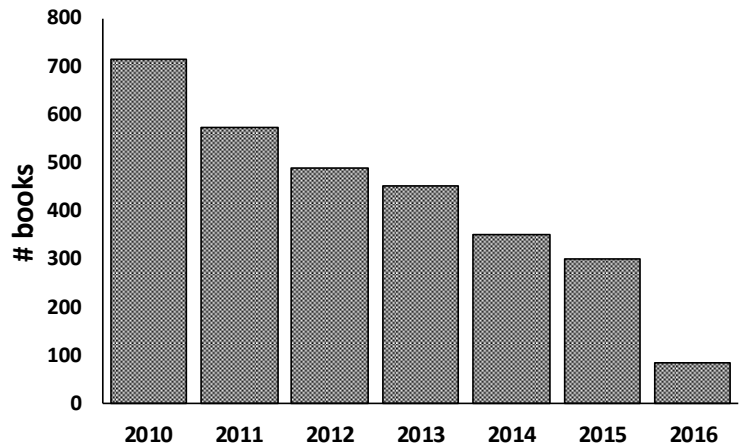

Figure 1. Number of monographs published by researchers from the University of Granada according to SICA in the 2010-2016 period

## Indicators used and PlumX as a data source

In this study, we analyse 18 indicators using three different proxies of impact: citations, downloads, library holdings and social media mentions. There are currently three big tools collecting and aggregating altmetric data: ImpactStory, Altmetric.com, and PlumX. Whereas altmetric.com and PlumX focus more on gathering and providing data at a large scale, ImpactStory's target group is the individual researcher who wants to include altmetric information in her CV (Peters et al.; 2016). Also, PlumX is the only one which provides other alternative metrics along with citation data, as well as social media mentions. For this study, we are using PlumX (the fee-based altmetrics dashboard). The data were gathered from PlumX in October 2, 2016. And they were permanently checked until the end of December 2016. No changes were reported in this period. However, further research might attempt to clarify the stability and reproducibility of altmetrics data, and provide thorough and transparent information regarding their temporal evolution and to trace and understand potential score changes. " PlumX uses ISBN numbers as book identifiers. Although no other alternative solution has been proposed so far, this imposes some limitations which should be noted. As indicated by Zuccala and Cornacchia (2016), multiple ISBN numbers may be assigned to the same 'work' due to the publication of new editions, translations or to its reprint and distribution by a different publisher.

| Category | Counts | Data Source | Type of Data Source |
|---|---|---|---|
| **Usage** | Abstract Views | Dspace | Repository |
| | Downloads | Dspace | Repository |
| | Sample Downloads | Ebsco | Electronic Provider |
| | Abstract Views | Ebsco | Electronic Provider |
| | Data Views | Ebsco | Electronic Provider |
| | PDF Views | Ebsco | Electronic Provider |
| | HTML Views | Ebsco | Electronic Provider |
| | Link outs | Ebsco | Electronic Provider |
| | Holdings | WorldCat | Libray Catalog |
| **Mentions** | Reviews | Amazon | Electronic Bookseller |
| | Reviews | Goodreads | -- |
| | Links | Wikipedia | Online reference |
| **Captures** | Export Saves | Ebsco | Electronic Provider |
| | Readers | Mendeley | Reference Manager |
| | Readers | Goodreads | Social platform |
| **Social Media** | Tweets | Twitter | Microbloggin Network |
| | Shares, Likes & Comments | Facebook | Social platform |
| **Citations** | Citation Counts | CrossRef | -- |

Tabla 1.  Data gathered in PlumX for this study according to their source of origin

Andrea Michalek and Mike Buschman founded Plum Analytics in early 2012. In 2014, it became Plum, a subsidiary of EBSCO Information Service[1].

Metrics are categorized in PlumX in five separate types: usage, captures, mentions, social Media, and citations. Table 1 includes the 18 indicators included in this study categorized by PlumX and their source of origin as well as the type of data source. Following we briefly discuss each of these categories:

**Usage.** This category includes abstract views, downloads, links-outs, library holdings and video plays from different data sources like DSpace, EBSCO and WorldCat.  As observed, two types of indicators are considered here, those related with electronic usage (downloads, views, etc.) and those related with usage in print format. The former has been traditionally referred to as *usage bibliometrics*, a concept coined by Kurtz & Bollen (2010) ad derived from the need to quantitatively assess the use of the electronic collections of libraries. The analysis of library holdings as a potential indicator for research assessment was originally envisioned by Torres-Salinas and Moed (2009) and White et al. (2010). In this case, the use of library holdings was specifically proposed to measure the dissemination of monographs.

**Mentions.** It includes blog posts, comments, reviews and links from different tools like Wikipedia, Goodreads or Amazon. What we find here are mainly altmetric indicators which focus on mentions from social media platforms. Altmetrics, term coined by Jason Priem in a tweet (Priem, 2010) are defined in a rather ambiguous way as any type of mention to scientific

---

literature in any type of social media platform. While the value of altmetrics in research evaluation is still largely questioned (Sugimoto et al., 2016; Wilsdon et al., 2015), it has been suggested that it could be a plausible means to measure broader forms of impact (Bornmann, 2014). In this case, most of the indicators in this section are social reviews, an indicator suggested as potentially relevant to assess the impact of books (Kousha & Thelwall, 2016).

**Captures.** This category, also includes altmetric indicators. In this case, we find bookmarks, code forks, favorites, readers and watchers from different tools like Mendeley, Goodreads or EBSCO. These are indicators usually related with readership metrics (Haustein, 2014). In the case of Goodreads, this data source has also been explored as to regard to its potential to assess the impact of books (Kousha, Thelwall & Abdoli, 2016; Zuccala et al., 2015).

**Social Media.** This category includes +1s, likes, shares, tweets from tools like Twitter or Facebook. These are altmetric indicators largely explored in the literature (especially with regard to Twitter), where no relation has been found with citations. While they cover a large portion of the journal literature, their relevance is still under question (Thelwall et al., 2013).

**Citations.** In this case, citations are retrieved from Cross-Ref, Scopus or patent and clinical citation data sources.

This categorization may be subject of criticism, but one its advantages is that the results are differentiated according to the indicator and their origin and can be aggregated according to the user's criteria.

Table 2 shows an example of the information retrieved from PlumX for two books included in our dataset.

| **Pereyra, M. A., Kotthoff, H. G., & Cowen, R. (2011).** PISA under examination. Sense Publishers. ISBN 9460917400 | | **García Godoy, María Teresa (2012).** El español del siglo XVIII: Cambios diacrónicos en el primer español moderno. *Peter Lang. ISBN* 3034310581 | |
|---|---|---|---|
| Captures:Exports-Saves:EBSCO | 0 | Captures:Exports-Saves:EBSCO | 32 |
| Captures:Readers:Mendeley | 9 | Captures:Readers:Mendeley | 0 |
| Captures:Readers:Goodreads | 0 | Captures:Readers:Goodreads | 0 |
| Citations:Citation Indexes:CrossRef | 0 | Citations:Citation Indexes:CrossRef | 1 |
| Social Media:Tweets:Twitter | 2 | Social Media:Tweets:Twitter | 0 |
| Social Media::Facebook | 2 | Social Media::Facebook | 0 |
| Mentions:Reviews:Amazon | 0 | Mentions:Reviews:Amazon | 0 |
| Mentions:Reviews:Goodreads | 0 | Mentions:Reviews:Goodreads | 0 |
| Mentions:Links:Wikipedia | 3 | Mentions:Links:Wikipedia | 0 |
| Usage:Sample Downloads:EBSCO | 0 | Usage:Sample Downloads:EBSCO | 0 |
| Usage:Abstract Views:DSpace | 0 | Usage:Abstract Views:DSpace | 0 |
| Usage:Abstract Views:EBSCO | 225 | Usage:Abstract Views:EBSCO | 335 |
| Usage:Data Views:EBSCO | 0 | Usage:Data Views:EBSCO | 0 |
| Usage:Holdings:WorldCat | 352 | Usage:Holdings:WorldCat | 842 |
| Usage:PDF Views:EBSCO | 0 | Usage:PDF Views:EBSCO | 131 |
| Usage:HTML Views:EBSCO | 0 | Usage:HTML Views:EBSCO | 42 |
| Usage:Downloads:DSpace | 0 | Usage:Downloads:DSpace | 0 |
| Usage:Link-outs:EBSCO | 11 | Usage:Link-outs:EBSCO | 3 |

Table 2. Example of indicators obtained from PlumX for two monographs published by researchers from the University of Granada

# Results

This section is structured in two parts. Each subsection is related with one of the specific objectives of the paper. First subsection analyses the coverage and distribution of the 18 indicators for the total production of books of the University of Granada during the 2010-2016 period. The second subsection compares the coverage by fields of citation indicators based on journal output with the coverage of PlumX indicators based on book output.

## Coverage and distribution of 18 impact indicators

2,299 books were identified in PlumX, representing 78% for our original dataset. However, 1382 books had no impact metric related to them. This represents 60% of the sample. Figure 2 shows for those which had metrics related to them, the distribution of indicators by metric category. As observed, 79% of the indicators are related with usage, followed by 20% of indicators related with captures. Significantly, mentions, citations and social media represent only 1% on the metrics identified. Within the usage category, the most predominant indicator is that related with library holdings obtained from the WorldCat catalogue (48%), followed by abstract views (22%). The low coverage for all indicators is reflected in table 2, where the indicator with the highest coverage (library holdings) only includes 31% of the total sample, followed by far by Mendeley readership (19%).

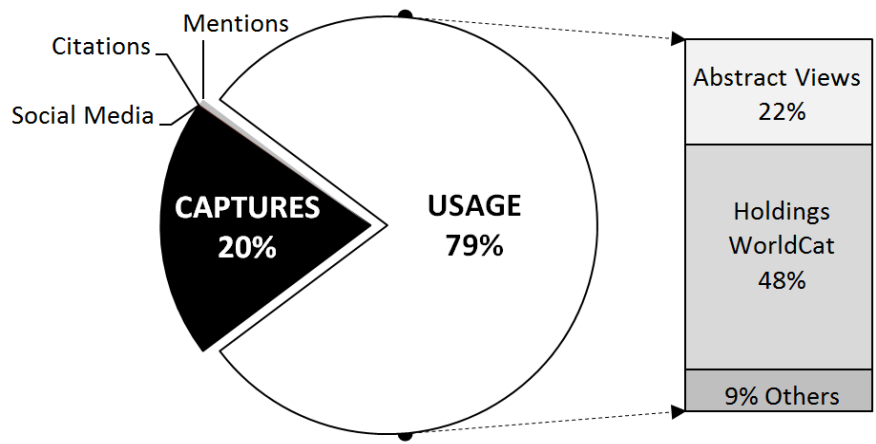

Figure 2. Distribution of indicators retrieved according to their category

| | # Books without metrics | % Books without metrics | METRICS | | | |
|---|---|---|---|---|---|---|
| | | | Sum | Avg by book | Std deviation | Max. value |
| Sample Downloads | 2,278 | 99% | 48 | 0.0 | 0.3 | 9 |
| Abstract Views (DSpace) | 2,298 | 100% | 237 | 0.1 | 5.0 | 237 |
| Abstract Views (EBSCO) | 1,752 | 76% | 21177 | 9.2 | 74.2 | 2,084 |
| Data Views | 2,298 | 100% | 7 | 0.0 | 0.2 | 7 |
| Holdings | 1,585 | 69% | 45545 | 19.8 | 103.1 | 1,271 |
| PDF Views | 2,257 | 98% | 2013 | 0.9 | 15.3 | 638 |
| HTML Views | 2,223 | 97% | 4278 | 1.9 | 22.2 | 782 |
| Downloads | 2,295 | 100% | 344 | 0.2 | 7.2 | 344 |
| Link-outs | 2,048 | 89% | 1384 | 0.6 | 5.7 | 244 |
| **Total usage** | **1,382** | **60%** | **75033** | **32.6** | **1574** | **4578** |
| Exports-Saves | 2,133 | 93% | 1794 | 0.8 | 7.0 | 197 |
| Readers (Mendeley) | 1,851 | 81% | 1423 | 0.6 | 2.5 | 41 |
| Readers (Goodreads) | 2,172 | 95% | 15722 | 6.9 | 212.0 | 9,716 |
| **Total captures** | **1,667** | **72%** | **18939** | **8.2** | **447.9** | **9718** |
| Tweets | 2,295 | 100% | 9 | 0.0 | 0.1 | 3 |
| Social Media | 2,297 | 100% | 11 | 0.0 | 0.2 | 9 |
| **Global Social Media** | **2,295** | **100%** | **20** | **0.0** | **0.5** | **11** |
| Reviews (Amazon) | 2,272 | 99% | 59 | 0.0 | 0.4 | 10 |
| Reviews (Goodreads) | 2,272 | 99% | 342 | 0.2 | 3.6 | 147 |
| Links | 2,280 | 99% | 27 | 0.0 | 0.2 | 4 |
| **Global Mentions** | **2,234** | **97%** | **428** | **0.2** | **9.6** | **147** |
| Citations (CrossRef) | *2285* | *100%* | 48 | 0,02 | 0,58 | 27 |
| **Global Citations** | ***2285*** | ***100%*** | **48** | **0,02** | **0,58** | **27** |

Table 2. Coverage and statistical indicators for metrics extracted from PlumX for the 2010-2016 period

In all, we observe that five of the indicators practically did not cover any of the records included in our sample (abstract views from DSpace, data views, downloads, tweets and social media), and seven covered less than 10% of the records (sample downloads, PDF views, HTML views, exports-saves, readers from Goodreads, reviews from Amazon and Goodreads, and links). The category with the lowest coverage is Social Media. Contrarily to what we observe with journal data (Robinson-Garcia et al., 2014), the coverage of Twitter is extremely low. Only four books include mentions in Twitter. Practically 100% of the records in our sample had no mentions in Twitter nor Facebook.

When focusing on the number of hits by book received for each indicator, we observe again, low figures on the average of metrics by book. Indeed, in 13 of the indicators used, the average number of hits is below one. However, we note considerable differences between the other indicators. Library holdings shows the highest average of hits by book (19.8) followed by abstract views in EBSCO and readers in Goodreads (6.9). The relation between the former and the latter has been suggested elsewhere (Zuccala et al., 2015) as an explanation for finding such a high average of readers despite its low coverage. In most cases, we also observe high deviation values, signifying the skewed distribution of these indicators, following the pattern of citation distributions. This is observed from the maximum number of hits reached by indicators. The largest number of hits for single books is found for readers in Goodreads with almost 10,000 hits, followed by abstract views in EBSCO (2,084) and library holdings (1,271).

The skewness of the distribution is confirmed by figure 3, where we analyse for the different categories of indicators the distribution of hits by the number of books. While all categories show a skewed distribution, the one with the lowest skewness is the category of usage.

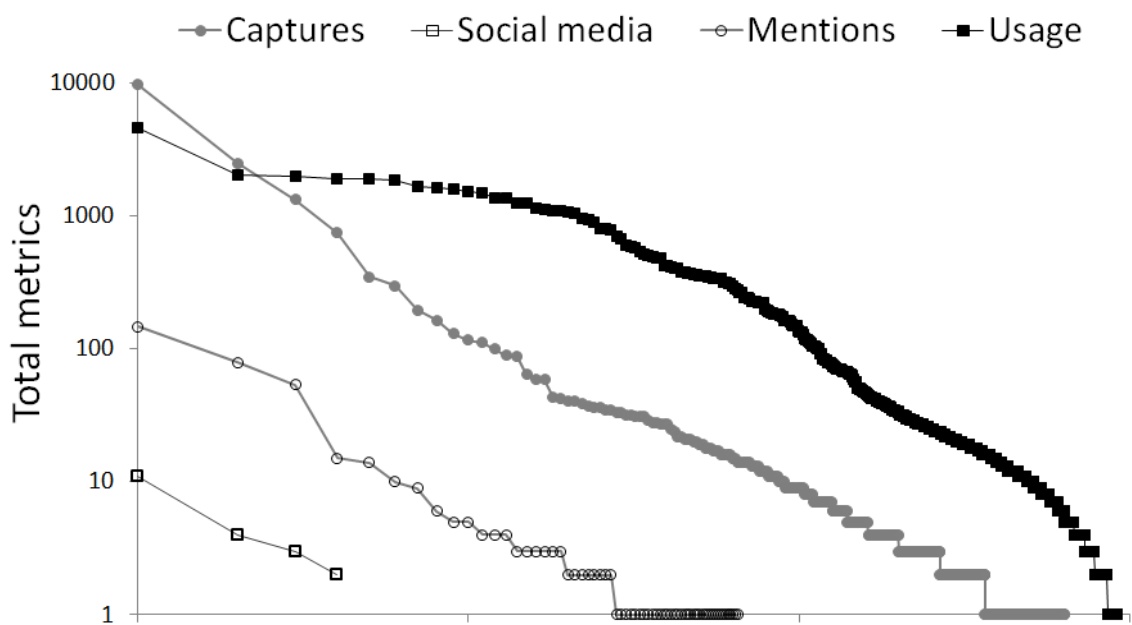

Figure 3. Distribution of hits by number of books according to the categories of indicators defined by PlumX for monographs published by researchers from the University of Granada in the 2010-2016 period

## Comparing a citation analysis of journal output vs. a multidimensional analysis of book output

At this stage, a key question is the extent to which the indicators offered by PlumX are useful. For this, figure 4 compares the output of the University of Granada and their impact depending on the document type and the impact indicators employed. Figures 4A and 4C show the university's output based on the number of published books and journal articles respectively. As observed, Humanities & Arts (1,175) and Social Sciences & Law (648) are the fields with the largest number of published books. Contrarily, when focused on journal articles, it is Natural & Exact Sciences (7,420) and Engineering & Technology (2,406). A similar pattern we observe in figures 4B and 4D. Indeed, figures 4C and 4D show the classical distribution of publications and citations based on Web of Science journals.

Clearly, the Humanities & Arts and Social Sciences & Law are the most negatively affected fields by bibliometric analyses: they have "little" output and "little" impact. It should be stressed that "little" is not used as a pejorative term but following the long tradition based on the famous book "Little Science, Big Science" by De Solla Price. Scientific publishing is a very complex activity and differs according to the different publication communities. Therefore, it is not possible to assess it by just counting together different publication outputs, like for example books and journal articles. It should be considered that the time involved in the writing and publishing of a book is much longer in comparison with a journal article. Concerning impact, the term "little" is even more disputable because there is a very strong dependence on the metric used or available as this study corroborates.

However, when focusing on alternative metrics such as the ones provided by PlumX, a completely different picture is shown. Figures 4A and 4B show and opposed view where these fields are the ones best represented. Combining both approaches we can provide a more accurate picture of the scientific impact and output, by introducing a neglected output (books) and more appropriate metrics to analyse their impact (e.g., library holdings). This avoids current mismatches in bibliometric analyses by broadening out the scope of outputs and opening up the type indicators used (Rafols, Ciarli, van Zwanenberg, & Stirling, 2012). Still, it should be noted that the indicators availability is limited to a reduced percentage of the output sample and, therefore, that complementariness is not always achieved.

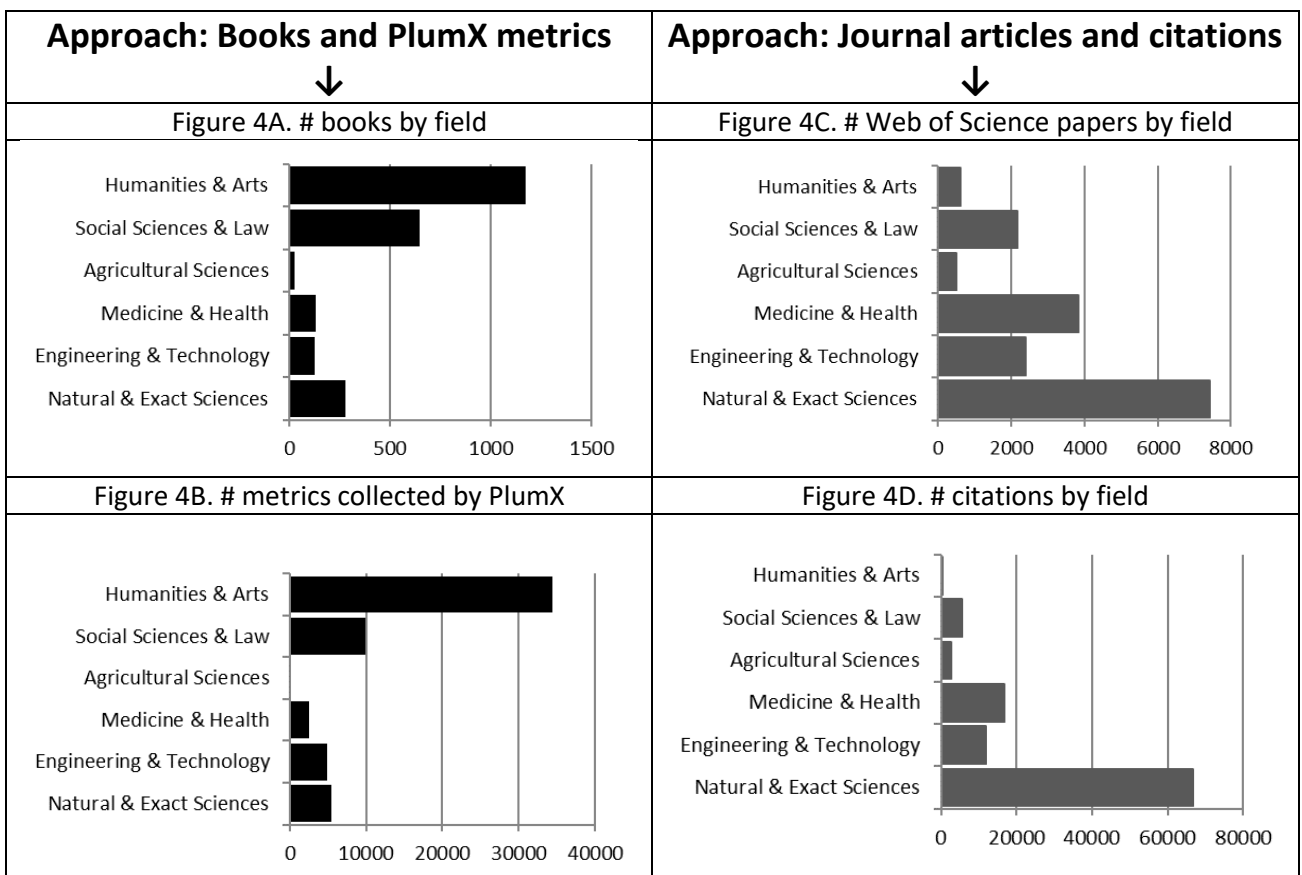

Figure 4. Comparative of approaches taken to analyse the scientific output of the University of Granada: Books vs. Journal articles. 2010-2016 period

## Discussion and concluding remarks

This study analyses the coverage and distribution of 18 indicators retrieved from PlumX of scholarly impact for books published by the University of Granada during the 2010-2016 period. These indicators are grouped into five categories, each aimed at showing different dimensions of impact. These are usage, mentions, captures, social media and citations. The aim of the paper is twofold. First, to analyse the coverage of PlumX indicators for monographs. Second, to compare traditional citation analyses based on journal articles with this multidimensional perspective offered by PlumX.

60% of the books included in our sample showed no values for any of the 18 indicators analysed. While this coverage may seem low, it is actually higher than that reported for citation. Torres-Salinas and colleagues (2014a) reported an uncitedness rate of 80.5% for the Book Citation Index. Usage indicators and specifically, library holdings were found to be the most comprehensive indicator for monographs.79% of books showed some values for this indicator. Contrarily, indicators such as mentions or social media and citations were almost lacking (see table 2). In this sense, it is worth mentioning the low figures found for tweets, an altmetric indicator which has been found to be the most widely-used data source for altmetrics (Thelwall et al., 2013). This could be related with the current crisis observed in the book publishing industry, where digital publishing has not expanded as much as with e-journals (Williams et al., 2009) and Open Access remains a challenge (Eve, 2014). This lack of mentions in social media could be due to the impossibility to access electronically to books.

Regarding the second goal. The comparison between approaches based on books and a variety of alternative indicators vs. those based on journal articles and citation-based indicators, shows once again, the importance of monographs in the social sciences and humanities. But also, the limitations of citation indicators to fully capture their impact. As pointed out in previous literature (e.g. Torres-Salinas & Moed, 2009; White et al., 2009), library holdings seem to be the most promising proxy of scholarly impact. Recent studies based on the Book Citation Index, show that citations are too scarce as to be considered as an appropriate impact measure for books (e.g., Torres-Salinas et al., 2014a,b). However, further research is still needed to surpass the many technical issues present when matching metadata from different sources with regards to monographs (Zuccala & Cornacchia, 2016).

The results shown here explore the potential interest on the variety of indicators offered by PlumX. But still, the results of this study rely very much on the features and abilities of PlumX. Many of them need to be studied in more detail, especially the ones concerning the correctness, validity and stability of the resulting data.

## Acknowledgments

The authors thank Stephan Buettgen (EBSCO) for granted trial access to PlumX. Nicolas Robinson-Garcia is currently supported by a Juan de la Cierva-Formacion grant from the Spanish Ministry of Economy and Competitiveness.